\documentclass[journal]{IEEEtran}

\usepackage{graphicx}
\usepackage{amsmath, nccmath}
\usepackage{amssymb}
\usepackage{afterpage}
\usepackage{verbatim}
\usepackage{psfrag}
\usepackage{algorithm}
\usepackage{lipsum}
\usepackage{color}
\usepackage{cite}
\usepackage{subfigure}
\usepackage{setspace}
\usepackage{epsfig}
\usepackage{epstopdf}
\usepackage{footmisc}
\usepackage{fmtcount}
\usepackage{stackrel}
\usepackage{float}
\usepackage{soul}
\usepackage{flushend}
\usepackage{enumerate}
\usepackage{color}
\usepackage{dblfloatfix}
\usepackage{tabularx,booktabs}
\usepackage{algorithmicx}

\begin{document}
\def\bm#1{\textbf{\em #1}}

\title{Beyond Line of Sight Defense Communication Systems: Recent Advances and Future Challenges}
\author{Ruhul Amin Khalil, \IEEEmembership{Member, IEEE}, Muhammad Haris, and Nasir Saeed, \IEEEmembership{Senior Member, IEEE}
\thanks{Ruhul Amin Khalil and Nasir Saeed are with the Department of Electrical and Communication Engineering, UAE University, Al-Ain 15551, UAE e-mail: ruhulamin@uaeu.ac.ae; mr.nasir.saeed@ieee.org.}
\thanks{Muhammad Haris is with the Department of Electrical Engineering, Hanyang University, Ansan 15588, South Korea (e-mail: haris737@hanyang.ac.kr).}

}

\maketitle
\begin{abstract}
Beyond Line of Sight (BLOS) communication stands as an indispensable element within defense communication strategies, facilitating information exchange in scenarios where traditional Line of Sight (LOS) methodologies encounter obstruction. This article delves into the forefront of technologies driving BLOS communication, emphasizing advanced systems like phantom networks, nanonetworks, aerial relays, and satellite-based defense communication. Moreover, we present a practical use case of UAV path planning using optimization techniques amidst radar-threat war zones that add concrete relevance, underscoring the tangible applications of BLOS defense communication systems. Additionally, we present several future research directions for BLOS communication in defense systems,  such as resilience enhancement, the integration of heterogeneous networks, management of contested spectrums, advancements in multimedia communication, adaptive methodologies, and the burgeoning domain of the Internet of Military Things (IoMT). This exploration of BLOS technologies and their applications lays the groundwork for synergistic collaboration between industry and academia, fostering innovation in defense communication paradigms. 
\end{abstract}

\maketitle

\IEEEdisplaynotcompsoctitleabstractindextext

\IEEEpeerreviewmaketitle

\section{Introduction}
\label{sec:introduction}
The global defense communication system market is poised for robust expansion, forecasted to reach a valuation of US \$46.7 billion in 2023 and expected to soar to an impressive US \$120.62 billion by 2033, exhibiting a substantial Compound Annual Growth Rate (CAGR) of 9.9\% \cite{defensemarket2023}. Comprising nearly a quarter of the global aerospace and defense market, this sector experiences heightened demand, owing to its reliance on commercial wireless technologies and an escalating need for satellite communication equipment. Regionally, North America commands market dominance with a 36.5\%  share in 2022, driven by substantial national defense expenditures. Europe, notably the United Kingdom, France, and Germany, account for a significant 24.8\%  market share. The Asia Pacific region, contributing 28\% in 2022, displays promising growth due to substantial government investments in military and defense equipment \cite{defensemarket2023}. While exhibiting slower growth, the Middle East and Africa are anticipated to show improvement by the forecast's end. Noteworthy efforts in modernizing defense communication systems include the United States' substantial investments in new handheld man-pack radios. With its robust industrial base, Germany stands prominently in Europe's military communication market. Simultaneously, China's marine communication systems industry foresees considerable expansion, fueled by coastal security demands and the growth of marine commerce operations \cite{defensemarket2023}. In a categorical breakdown, the sale of defense communication satellites exhibits a notable uptick, propelled by escalating tensions along international borders. The market's future trajectory hinges on strategic collaborations, continuous technological innovations, and the integration of the Internet of Things (IoT) \cite{gotarane2019iot}. Manufacturers have substantial opportunities for scaling their businesses by harnessing advanced communication systems, low-latency wireless links, and IoT integration, presenting a dynamic landscape for growth in the defense communication sector.

\begin{figure*}[h!]
\begin{center}
\includegraphics[width=0.9\textwidth]{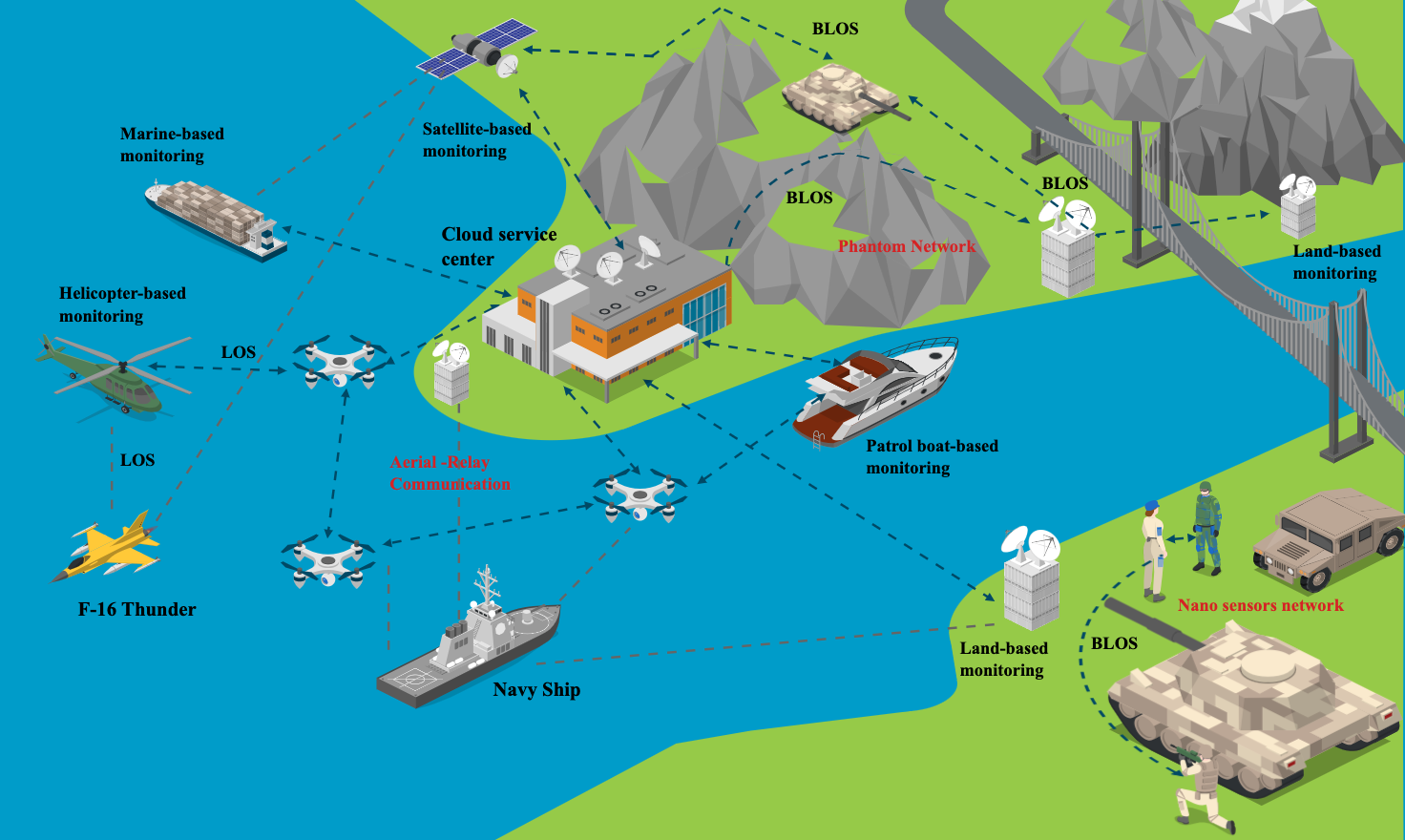}  
\caption{Illustraion of a BLOS defense communication system.\label{Fig1}} 
\end{center}  
\end{figure*}

Communication systems are a linchpin within military operations, enabling seamless information exchange among diverse units within the armed forces. Effective communication transcends mere necessity in the intricate threatening of defense; it assumes the mantle of a strategic imperative, wielding profound influence over the success of tactical missions and the overarching operational readiness \cite{rupar2020airborne}. This specialized communication domain grapples with challenges far beyond those encountered in civilian networks, confronting distinctive obstacles such as Electronic Warfare (EW), Command and Control Warfare (C2W), and the difficulties inherent in beyond-line-of-sight (BLOS) communication, as depicted in Figure \ref{Fig1}. The complexities inherent in defense communication amplify its strategic significance as the core enabling coordinated military action in the face of evolving threats and operational landscapes. Moreover, the multifaceted nature of military forces, spanning Ground, Water, and Air domains, introduces an additional stratum of complexity to defense communication. Serving as the sinew-binding coordinated military action, it equips forces to adapt, strategize, and prevail in the face of diverse, ever-evolving threats \cite{he2021wireless}. Defense communication, thus, emerges as the quintessential backbone, empowering troops to navigate the complexities of modern warfare and emerge triumphant amidst the dynamism of global defense landscapes.

Recently, five significant trends have reshaped the landscape of defense communications. First, space technologies are elevated in defense operations, where space-based connectivity is pivotal for various defense applications \cite{Saeed2020,khalil2020toward}. The military's investments prioritize purpose-built and resilient space systems functioning across diverse transports and orbits, envisioning space as a potential frontline in emerging conflicts. Secondly, the U.S. Department of Defense (DoD) increasingly embraces multi-constellation, multi-transport, and software-defined networks, with a focus on low earth orbit (LEO) systems for their resilience and low latency \cite{ricklober2022}. A future outlook envisions a fusion of LEO and geostationary satellites orchestrated via software-defined networking. Thirdly, pivotal efforts center around developing Joint All-Domain Command and Control (JADC), Project Convergence, and Project Overmatch. The objective is establishing a unified command and control system spanning diverse military branches, emphasizing collaboration with industry stakeholders. The fourth trend underscores the escalating reliance on commercial technology within defense networks, integrating commercial innovations and off-the-shelf products into space-based networks to bolster resilience \cite{ricklober2022}. Finally, standalone 5G deployments tailored for the DoD offer high-speed, secure, and private wireless communication at facilities, featuring a zero-trust security architecture and the capability to run edge applications employing artificial intelligence and machine learning. These trends align with strategic objectives to address end-user needs, foster collaboration, and explore commercial technologies to enhance defense communication capabilities.

While defense communication has attracted considerable scholarly focus, investigations span cognitive radios \cite{tuukkanen2021systematic}, blockchain-based privacy preservation \cite{wu2021exploiting}, dynamic spectrum anti-jamming communication \cite{wang2020dynamic}, Internet of Radars, joint Radar communication \cite{akan2020internet}, and BLOS phantom networks \cite{misra2020phantom}. However, to the best of the author's knowledge, a comprehensive examination of BLOS communication within defense communication systems is absent in existing literature. Therefore, this paper aims to bridge this gap by delving into state-of-the-art technologies and strategies explicitly tailored for BLOS scenarios in military contexts.

The remainder of this paper is organized as follows. First, we present state-of-the-art technologies in BLOS defense communication, encompassing phantom networks, nanonetworks, satellite networks, and aerial relays. Following this, a comprehensive case study illustrating UAVs for BLOS route planning in defense communication is presented. Subsequently, emerging challenges and future research directions are delineated, culminating in the conclusive section of this article.

\section{Cutting Edge Technologies for BLOS Defense Communication Systems}
Defense communication encompasses critical components such as military radios, C2W, EW, secure transmission, and alert measurement systems. Direct wave propagation, reliant on an unobstructed line of sight (LOS) between the Source (Transmitter) and Destination (Receiver), stands as an optimal solution for information exchange where frequencies below approximately 50 MHz favor surface or ground wave propagation \cite{poisel2008introduction}. Nevertheless,  higher frequencies face significant attenuation, necessitating ducting for signal propagation across substantial distances within two Troposphere layers. Moreover, LOS communication effectiveness can be hindered by obstacles like the Earth's curvature and natural or man-made impediments, requiring the adoption of BLOS communication techniques \cite{rupar2020airborne}. Military domains extensively rely on various BLOS techniques, including phantom networks, nanonetworks, satellite networks, and aerial relays.
Phantom networks provide an on-demand aerial communication method among out-of-range devices, while nanonetworks comprise nano nodes within atomized aqueous mists, forming the basis of phantom networks. Historically, balloons, drones, and unmanned aerial vehicles (UAVs) facilitate BLOS communication. Despite their utility, these BLOS techniques encounter challenges. Table \ref{table:tab1} offers a comparative overview of BLOS techniques, highlighting key features and associated challenges.
\begin{table*}[ht!] 
\newcolumntype{C}{>{\arraybackslash}X}
\setlength{\extrarowheight}{1pt}
\caption{Various BLOS communication networks and challenges.\label{table:tab1}}
\begin{tabularx}{\textwidth}{|C|C|C|}
\hline
\hline
\textbf{BLOS Networks} & \textbf{Applications} & \textbf{Challenges}  \\    
\hline
\hline
\textbf{Phantom Networks} & Facilitates seamless communication between military units in urban environments with complex structures to navigate obstacles in BLOS scenarios & Dynamic topology management with rapidly changing terrains and scenarios, such as urban environments, where network topology needs continuous adjustments for optimal communication\\
\hline
\textbf{Nanonetworks} & Provides covert surveillance in hostile BLOS environments, enabling discreet information gathering without compromising operational security & Limited energy  to sustain continuous communication in resource-limited BLOS environments \\
\hline
\textbf{Satellite-based Networks}  & Supports worldwide military operations by ensuring continuous communication and data exchange, even in remote or contested areas & High time delay  due to large transmission distance, impacting real-time communication in critical military operations  \\
\hline
\textbf{Aerial Relays}  & Provides rapid deployment scenarios for quick and flexible communication setups and provides BLOS connectivity during fast-paced operations & Suffers from communication disruptions caused by atmospheric interference or potential vulnerability due to electronic warfare \\
\hline
\hline
\end{tabularx}
\end{table*}


\subsection{Phantom Networks}
Phantom networks represent an innovative cornerstone in BLOS communication strategies for defense and military applications, signaling a paradigm shift with on-demand, intangible aerial connectivity crucial where traditional LOS communication falls short \cite{misra2020phantom}. Their adaptability and responsiveness prove indispensable in dynamic and challenging military environments, fortifying communication effectiveness and enhancing the versatility of military radio systems. For instance, phantom MESH combines state-of-the-art communication and advanced jamming technologies, employing MESH radios, a token passing-based mobile ad-hoc network \cite{andrew2023pmesh}. This self-healing network architecture establishes robust RF paths in challenging environments, crucial for operations beyond direct visual or radio contact. The iMESH suite within this solution, including iMESH KRIP and iHIVE, ensures flexibility, security, and rapid deployability. Other phantom technologies, like the Eagle 108 Drone Jammer and EW1600 military tactical satellite communication (SATCOM) jammer, neutralize threats from unauthorized drones, ensuring safety and security in BLOS scenarios. Blu Wireless introduces PhantomBlu, a groundbreaking V-Band radio node optimized for BLOS military communications, featuring dynamic mesh networking capabilities with multi-gigabit data rates suitable for diverse BLOS scenarios \cite{satnews2023pblu}. This system efficiently processes high-bandwidth sensor data and video at the tactical edge, providing covert and resilient communication for ground forces operating in BLOS environments. PhantomBlu's ongoing advancements aim to include W-band transceivers, aligning it with IEEE 6G standards and positioning it as a versatile BLOS solution for near-peer tactical operations.

\subsection{Nanonetworks}
Nanonetworks constitute a pioneering technological foundation in BLOS communication for defense and military applications \cite{akyildiz2008nanonetworks}. Operating at the nanoscale, these networks employ nano nodes suspended within atomized aqueous mists, providing essential sophistication for robust BLOS communication. Their intricacies offer a nuanced solution for surmounting environmental obstacles in dynamic operational theatres \cite{akyildiz2012monaco}. Their nanoscale components significantly enhance flexibility and adaptability, enabling effective communication in scenarios where traditional methods falter \cite{dressler2012towards}. Moreover, nanotechnology's potential impact on BLOS military applications extends to medical care and materials advancements. Specifically tailored for BLOS scenarios, military nanotechnology focuses on improving body Armor effectiveness. Technologies such as Silica $S_i$, Titanium dioxide $T_{i}O_{2}$, and Silicon dioxide $S_{i}O_{2}$ nanoparticles. These nanoparticles form the nanodevices that aim to offer soldiers lightweight yet resilient protection in situations where direct visual or radio contact is obstructed \cite{willsoutter2012}.
Furthermore, nanotechnology advances sensors, presenting smaller, more sensitive alternatives critical for BLOS operations. Examples include highly sensitive infrared thermal sensors, compact accelerometers with GPS for motion and position sensing, miniature high-performance camera systems, and biochemical sensors, illustrating the adaptability of nanomaterials in augmenting military sensor capabilities in BLOS environments. Integrating nanotechnology into BLOS robotic systems and control mechanisms further enhances efficiency, emphasizing nanotechnology's transformative influence on BLOS military technologies.

\subsection{Satellites-based BLOS Communication}
Safeguarding communication across vast and challenging terrains, satellite-based defense communication stands as a linchpin in BLOS operations, ensuring continuous connectivity during critical military missions and protecting sensitive data through encrypted transmission \cite{ussf2020,haris2021polarization}. However, persistent challenges like signal latency and susceptibility to electronic warfare underscore the need for ongoing technological advancements.
Key innovations, including emerging LEO constellations and adaptive communication protocols, are poised to augment the resilience and effectiveness of modern military communication strategies \cite{9466942}. The Defense Satellite Communication System (DSCS), serving as a critical component of global satellite communications for the U.S. military, comprises high-capacity military satellites supporting diverse defense entities with its six-channel transponder system \cite{defenseone2018}.
This constellation, known for its secure, nuclear-hardened, anti-jam, and high-data-rate communication, is integral for long-haul communications in contested environments. Other satellite-based defense communication networks such as SATCOMBw in Germany, Syracuse IV in France, and Skynet in the U.K. serve as critical communication relays in BLOS operations, ensuring uninterrupted connectivity in military scenarios \cite{miltsatcom2023}. These systems employ encrypted data transmission protocols, ensuring the confidentiality and integrity of defense communications. Ongoing advancements are targeted to address challenges such as signal latency and susceptibility to electronic warfare, aiming to bolster the effectiveness and resilience of modern military communication strategies.

\subsection{Aerial Relays}
This BLOS communication technology involves both manned aircraft and UAVs serving as airborne relays, establishing and maintaining communication links over extended distances \cite{yin2023air}. Manned aircraft with advanced communication systems act as crucial nodes in military communication networks, enhancing versatility and coverage in BLOS communication.
Meanwhile, UAVs present a dynamic and adaptable solution, adept at navigating challenging environments and swiftly deploying to establish communication links in contested airspace \cite{9832657}. Using tethered drones, tethered balloons, and floating balloons further diversifies the aerial relay landscape, offering unique solutions for sustained communication in diverse operational scenarios \cite{kim2018drone}. These aerial relays strategically overcome LOS limitations, ensuring continuous and resilient communication support for military operations. The following discusses prospective aerial relays for BLOS communication in defense communication systems.

\subsubsection{\textbf{Manned Aircrafts:}}
Manned aircraft are pivotal and multifaceted in BLOS communication within defense and military applications. Equipped with advanced communication systems, these aircraft act as airborne relays, forming crucial links over extended distances. Their versatility goes beyond conventional communication, enabling strategic deployment across diverse operational scenarios. These aircraft offer unparalleled coverage and adaptability, functioning as vital nodes within military communication networks. Beyond their role as relays, they contribute significantly to intelligence, surveillance, and reconnaissance (ISR) efforts, elevating the overall situational awareness of military operations. The human-operated nature of manned aircraft introduces dynamic decision-making, allowing real-time adjustments to communication strategies in response to the evolving needs of complex military environments. Manned aircraft are effective conduits for BLOS communication and stand as cornerstones in comprehensively integrating communication, surveillance, and strategic capabilities within defense operations.

\subsubsection{\textbf{Unmanned Aerial Vehicles (UAVs):}}
Heralded as transformative assets in modern military strategies, UAVs epitomize adaptability and innovation in BLOS communication. These unmanned systems, purpose-built for navigating challenging environments, contribute significantly to the military's communication capabilities where traditional methods prove impractical. UAVs bring unparalleled flexibility to BLOS scenarios, rapidly deploying to establish communication links in contested airspace, inaccessible terrains, or high-risk zones \cite{10077453, khalil2023uavs}. Beyond their role as communication relays, UAVs contribute extensively to intelligence gathering and surveillance, enhancing situational awareness in dynamic military landscapes. The autonomous nature of UAVs allows for swift and agile responses to changing operational demands, showcasing their pivotal role in real-time decision-making processes. As integral components of military BLOS strategies, UAVs bolster communication resilience and embody a transformative force in enhancing the agility and effectiveness of defense operations in an evolving and complex security landscape.

\subsubsection{\textbf{Tethered Drones:}}
Tethered Drones emerge as a distinctive and versatile solution within BLOS communication for defense and military applications. These drones, tethered to a fixed point, present a unique capability for sustained communication, making them suitable for prolonged surveillance and communication relay tasks. Tethering enables prolonged flight duration and ensures a stable platform, contributing to persistent and reliable communication support. In military scenarios, tethered drones offer adaptability in various operational contexts, providing a dynamic vantage point for surveillance and communication even in challenging environments. Their ability to hover at specific altitudes for extended periods makes them valuable assets for maintaining continuous communication links in complex, dynamic military theatres. Beyond their role in communication, tethered drones also contribute to intelligence gathering, surveillance, and monitoring, showcasing their multifaceted utility in enhancing military capabilities beyond BLOS communication resilience. The strategic use of tethered drones underscores their significance as agile and enduring tools in modern defense communication strategies.

\subsubsection{\textbf{Tethered Balloons:}}
Tethered Balloons present a distinctive and strategically valuable solution within BLOS communication for defense and military applications. Tethered securely to the ground, these balloons serve as stable, elevated platforms for improved signal propagation. Tethering provides a cost-effective solution and ensures enduring and consistent communication links over expansive areas. In military applications, tethered balloons can be strategically positioned to enhance coverage and connectivity in specific operational zones, offering a dynamic and adaptable means of establishing communication in challenging terrains. Their stability, adaptability, and capacity to operate at various altitudes make them valuable assets for maintaining communication resilience in diverse and dynamic military environments. Beyond their role in communication, tethered balloons contribute to intelligence, surveillance, and reconnaissance efforts, showcasing their versatility as integral components of military operations. With their enduring presence and strategic positioning, the tethered balloons exemplify a nuanced approach to enhancing BLOS communication capabilities, providing a reliable and comprehensive solution in modern defense scenarios.

\subsubsection{\textbf{Floating Balloons:}}
Floating Balloons, equipped with communication payloads, represent an innovative and dynamic solution within the realm of BLOS communication for defense and military applications. These balloons traverse the skies at high altitudes, offering a mobile and adaptable platform for maintaining communication links in challenging terrains. Their mobility and ability to operate at varying altitudes render floating balloons effective tools for ensuring continuous communication coverage across diverse operational scenarios. In military applications, these balloons contribute significantly to establishing resilient and expansive BLOS communication networks, particularly in remote or inaccessible areas where traditional methods may fall short. The innovative approach of floating balloons aligns with the evolving demands of modern warfare, offering a strategic advantage by providing flexible and robust communication capabilities. Beyond their role in communication, floating balloons contribute to intelligence gathering, reconnaissance, and surveillance efforts, showcasing their multifaceted utility in enhancing military capabilities beyond BLOS communication resilience. The strategic deployment of floating balloons underscores their significance as forward-looking and versatile assets in the ever-evolving landscape of defense communication strategies.

\section{Case Study: UAVs for BLOS Route 
 Planning in Defense Communication Systems}
As discussed in the previous section, in defense communication systems, UAVs play a crucial role in addressing BLOS challenges where various scenarios are marked by obstructed direct visual or radio contact, demanding innovative solutions for establishing robust communication links. As agile and adaptable platforms, UAVs navigate complex terrains and operate amidst adversarial threats, overcoming hurdles in BLOS environments \cite{wang2019completion}.
Strategic deployment of UAVs for BLOS route planning becomes essential, ensuring optimized paths through intricate operational environments to enhance communication reliability and mission success. Integrating cutting-edge technologies like path optimization techniques into UAV-based BLOS route planning can further enhance the efficacy \cite{bollino2008collision}. Path planning for UAVs in challenging environments presents a real-world optimization challenge, requiring an optimal planner for efficient navigation. It involves conceptualizing path-planning missions as optimization problems, where evolutionary computation emerges as an efficient method for addressing such challenges. Nature-inspired optimization techniques such as particle swarm optimization (PSO) can dynamically optimize UAV flight paths, which is crucial in navigating radar-laden war zones. This integration fortifies defense communication systems, enhancing agility, responsiveness, and security, enabling seamless communication links across challenging distances, significantly contributing to mission success \cite{dinc2009design}. This section focuses on UAV path planning in radar-threat war zones, framing it as an optimization problem.

\begin{figure}[h]
\begin{center}
\includegraphics[width=0.5\textwidth]{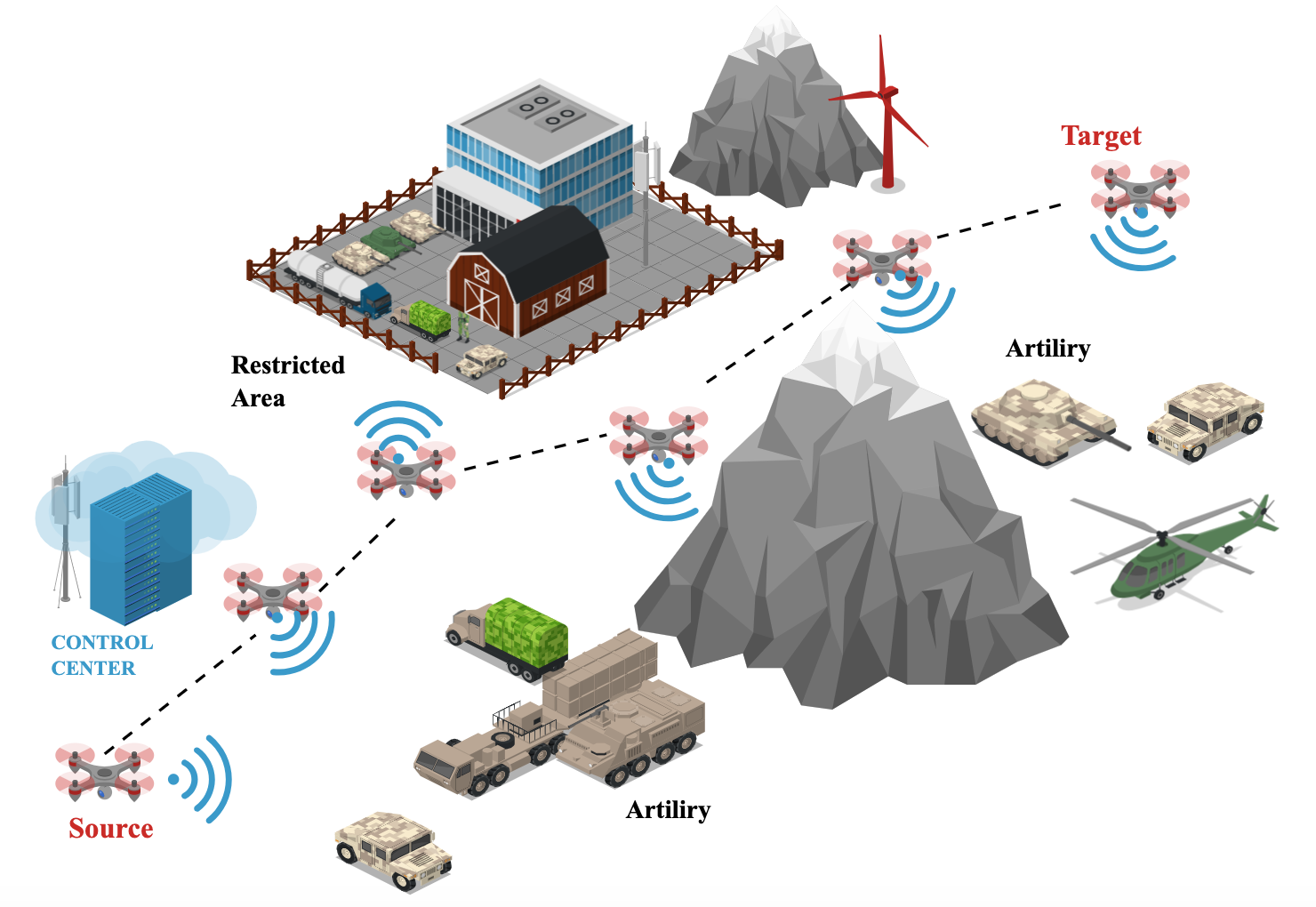}
\caption{An Illustration of UAVs path planning in military operations.}\label{UAVswarm}
\end{center} 
\end{figure}

We focus on path planning for a UAV within a challenging radar-limited environment, where deterministic search algorithms prove impractical due to vast search space complexities. The presence of regularly positioned circular radar threats, confined within predetermined global limits, further complicates the optimization process. The approach prioritizes optimizing UAV performance within a predefined 3D terrain environment. Efficient path planning for UAVs in military operations, between start and target points, is crucial for reduced time and enhanced mission effectiveness. The planning process involves three interlinked phases: (1) establishing cost and constraint functions, (2) aligning the digital map with the mission environment, and (3) optimizing and refining the flight route. In combat scenarios, UAV navigation encounters challenges from diverse danger sources such as topography, radars, artillery, and no-fly zones. Considering the UAV's physical constraints, the planned route must navigate these obstacles to ensure safe traversal and collision avoidance. Addressing the multitude of objectives and constraints arising from various risks and limitations presents challenges in UAV route planning \cite{path2022}. Developing an effective planner is crucial to handling the complexities of multi-objective limited UAV route planning, particularly in intricate situations, as shown in Fig. \ref{UAVswarm}.

In UAV path planning, PSO significantly optimizes each particle's path within predefined constraints and initial velocities. Despite their inclination to move towards their velocity vectors, particles in PSO abide by swarm constraints, exchanging information and retaining their previous best position alongside their current position and velocity. The swarm's overall best particle, known as the global best, emerges in PSO iterations. PSO's iterative process refines particle positions by updating their location and velocity \cite{SHPSO}. During path optimization, particles use vectors like their previous velocity, the global best position, and the personal best position to determine optimal routes. In the context of UAV path planning, this optimization method guides UAVs through complex environments, facilitating efficient navigation towards mission objectives. Leveraging PSO's ability to find optimal solutions within challenging and dynamic scenarios, this approach assists UAVs in identifying safe and effective routes amidst diverse obstacles and constraints.

\begin{figure*}
   \centering
   \subfigure[]{\includegraphics[scale=0.40]{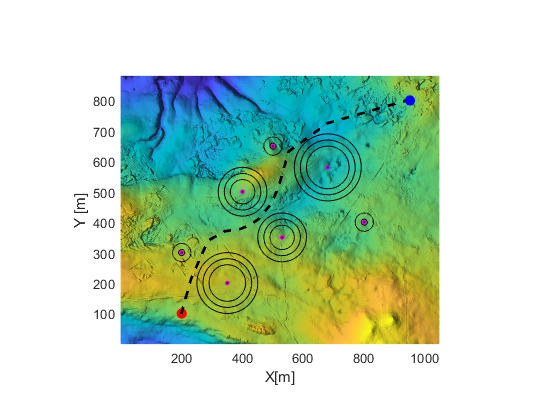}} 
   \subfigure[]{\includegraphics[scale=0.40]{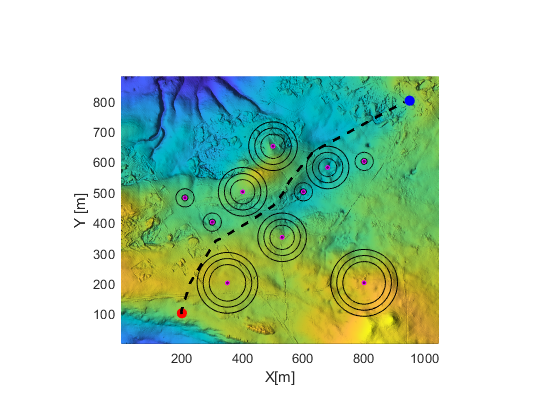}} 
   \subfigure[]{\includegraphics[scale=0.40]{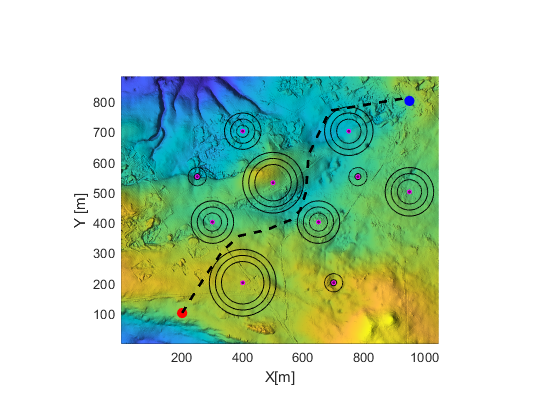}}
   \caption{ UAV route planning in different scenarios (a)  low complexity (b) medium complexity (c) high complexity. }
   \label{s1}
\end{figure*}

\begin{figure}
\centering
\includegraphics[scale=0.64]{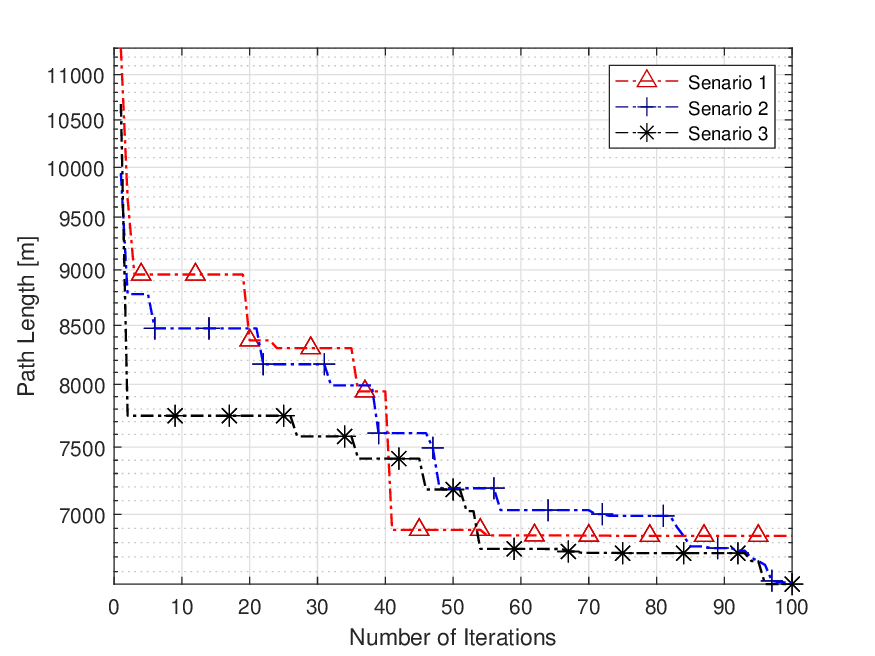}
\caption{Convergence curve of 3 different scenarios}
\label{r1}
\end{figure}

Utilizing the PSO algorithm, we formulated optimal flight paths for UAVs navigating through combat zones, particularly in the presence of radar systems and ground-based artillery. Radar systems are critical in locating and tracking airspace objects, making evading radar detection imperative for mission success and security in hostile environments. Additionally, ground-based artillery poses a significant threat to UAVs. Addressing these ground-based threats requires navigation strategies considering not just altitude but also the dynamic nature of the threats, demanding agile and adaptable flight paths. Effectively evading both hostile radar detection and ground-based threats, like artillery, requires a comprehensive strategy that recognizes the distinct challenges posed by each threat source.
Our proposed UAV's trajectory routing is designed to maintain a consistent altitude above sea level, simplifying the navigation model and establishing a reliable benchmark for the UAV's flight trajectory. In contrast, ground-based navigation often involves adaptations to the terrain and on-ground obstacles, necessitating deviations from the UAV's consistent altitude approach. The UAV's route planning based on the PSO algorithm is depicted in Fig. \ref{s1}, where a small red circle represents the initial UAV position, while a blue circle denotes the target location. Large circles indicate radar threats, while small circles represent artillery threats. The black dotted line illustrates the path generated by the PSO algorithm. The scenarios varied in complexity, giving unique navigation problems.  We assume the UAV has prior knowledge of the target location and aims to navigate while evading various threats to reach its designated objective.
Three scenarios with varying complexity levels have been introduced, each presenting different configurations of radar and artillery threats. Higher complexity levels entail more intricate radar and artillery threats distributed sparsely along the trajectory from the starting point to the goal. We also illustrate the convergence rate concerning the path length for all three scenarios in Fig. \ref{r1}.

\section{Challenges and Future Research Directions} 
 The evolving landscape of modern warfare demands communication systems adept at maneuvering through diverse networks, managing contested electromagnetic spectrums, and meeting escalating requirements for multimedia-intensive communications. Integrating unmanned autonomous systems and diverse sensors in military operations elevates the importance of adaptive communication strategies, particularly in BLOS scenarios. This section delves into the cutting-edge challenges unique to BLOS contexts, offering valuable insights into the future of military communication technologies and their pivotal role in shaping successful defense operations, especially under conditions where direct visual or radio contact is limited.

\subsection{Resilience in Harsh Infrastructure Environments}
Defense communication systems, operating across diverse and dispersed environments, encounter infrastructure challenges distinct from those in established commercial networks \cite{thales2023}. The adaptability of defense communication systems becomes paramount for ensuring reliable and secure connectivity in unpredictable operational scenarios, spanning various terrains, from dense urban landscapes to remote, austere regions. Effectively addressing these BLOS infrastructure challenges underscores the need for strategic approaches to optimize network performance and enhance resilience across dynamic landscapes. An emerging frontier for BLOS research involves exploring the evolving role of space-based systems within critical infrastructure frameworks, offering pivotal opportunities for future military operations \cite{coman2019critical}. Understanding and harnessing the potential of BLOS space-based systems could revolutionize the planning and execution of military endeavors. The reliability of these systems, integral to most BLOS military operations, highlights the interdependence between BLOS satellite control and underground systems \cite{coman2019critical}. Future research should explore the intricate relationship between national security and safeguarding critical infrastructure. Considering the persistent evolution of threats driven by technological advancements, innovative approaches, including adaptations to federal regulations and proactive anticipation of potential crisis scenarios, should be explored to fortify this essential connection.

\subsection{Heterogeneous Networks for BLOS Communication}
Heterogeneous networks are crucial in BLOS defense communication systems, offering a unified infrastructure that seamlessly integrates diverse technologies, standards, and devices \cite{chen2020joint}. This approach is instrumental in tackling challenges arising from temporal and spatial variability during military operations, providing adaptive solutions that dynamically adjust to changing conditions, and ensuring continuous communication \cite{singh2021threshold,suri2016realistic}.
The increasing integration of Commercial Off-The-Shelf (COTS) technologies in military communications, driven by advantages like simplified procurement, rapid deployment, and cost-effectiveness, necessitates meticulous customization to adhere to stringent security and reliability standards. Tactical communications, marked by the coexistence of COTS-derived solutions and military standards, require precise planning and integration testing for heterogeneous networks in BLOS communication scenarios. For instance, the Italian Navy's research center, Selex ES, and ELMAN performed a collaborative effort to develop a modular testbed that facilitates swift integration of diverse technologies, providing a customizable environment for various test campaigns, particularly relevant to maritime and littoral operations \cite{arreghini2015heterogeneous}. These initiatives highlight the broad applications of heterogeneous networks in military contexts, addressing challenges in experimentation, security, routing architectures, energy efficiency in sensor networks, and optimization of UAV coalition-based networks for surveillance and spectrum access \cite{hegland2020federating}. Together, these efforts underscore the pivotal role of heterogeneous networks in meeting the intricate communication requirements of modern military operations in BLOS scenarios.

\subsection{Contested Electromagnetic Spectrum}
Military communications encounter the challenges of navigating contested electromagnetic environments, where adaptability remains pivotal for sustaining BLOS link performance and resilience against vulnerabilities \cite{defense201521st,hoehn2020overview}. 
BLOS considerations gain significance as military operations extend beyond direct visual or radio contact. This strategic shift aims to secure freedom of action in the contested electromagnetic spectrum while recognizing the limitations of BLOS in seeking global supremacy through rapid technological advancements. In the contemporary global landscape, characterized by widespread access to high-end electronics technology, the Defense Science Board conducted a comprehensive investigation into the U.S. military's operational capabilities within complex electromagnetic environments \cite{bahk2020traversing}. This investigation revealed significant deficiencies in operational support across diverse mission areas, highlighting potential limitations against near-peer and regional powers in terms of sensing, communication, networking, and synchronizing operations. The electromagnetic spectrum has emerged as a strategic domain, interlinking land, air, naval, space, and cyber domains. While Western armed forces historically maintained electromagnetic superiority, the evolving landscape challenges this dominance with its increasingly complex and congested spectrum. Competitors such as Russia, China, and non-state actors leverage advancing technology, threatening Western sovereignty in this domain. Addressing this challenge requires a pragmatic approach emphasizing issues like agile spectrum management, effects concentration, and subsidiarity at the tactical level to establish electromagnetic local superiority within specific space-time frameworks.

\subsection{Increasing Need for Multimedia-Intensive Communication}
In contemporary military operations, the scope of BLOS communication surpasses traditional voice and data transmission to encompass highly multimedia-intensive capabilities, including video and broadband. This evolution mirrors the dynamic nature of 21st-century military competencies, where soldiers are required to embody essential attributes (Be), possess knowledge (Know), and execute actions (Do) for success in diverse operational environments \cite{sangwan2021philosophy}.
Addressing the challenges of flexible, rapidly deployable communication systems within the broader spectrum of BLOS military communication also acknowledges the need for multimedia-intensive capabilities. The Defense Advanced Research Projects Agency (DARPA) underscores the imperative of untethered communications, advocating for research and development efforts to meet the Department of Defense's requirements for security, interoperability, and flexibility \cite{national1998evolution}. 5G and beyond technologies can be investigated to provide real-time decision support in the future defense landscape, enabling hyper-converged connectivity and secure data networks. Moreover, it can facilitate the transfer of massive amounts of data, providing instant situational awareness and enhancing training and battlefield capabilities.

\subsection{Adapting to Changing Requirements}
Integrating unmanned autonomous systems and diverse sensors poses dynamic and evolving requirements, demanding high agility and adaptability from military communication systems. These evolving needs are pivotal for ensuring effective connectivity in BLOS communication amid rapidly changing operational landscapes. While military establishments were once a smaller part of the wireless communications market, the current commercial sector significantly shapes technology development \cite{vassiliou2013crucial}.
However, crucial disparities persist, including the necessity for BLOS communication in areas lacking well-maintained infrastructure, the integration of multi-hop networks, and the operation of multiple heterogeneous networks within the same area. Bridging these diverse networks, especially in contested electromagnetic environments, remains a formidable challenge \cite{knox2018socio}.
As the military grapples with evolving operational landscapes, fostering a new ethic for analysis and analysts and emphasizing cognitive engineering principles becomes paramount in ensuring the adaptability and effectiveness of BLOS military communication strategies.

\subsection{Towards the Internet of Military Things (IoMT)}
In the realm of defense systems, the future lies in a unified network powered by the Internet of Military Things (IoMT). This interconnected system brings together various assets—ships, aircraft, ground vehicles, drones, and personnel—enabled by edge computing, AI, and 5G technologies. It enhances command structures, allowing instant decisions in real-time scenarios \cite{langleite2021military}.
BLOS communication is pivotal in shaping the future of IoMT, which unifies three key domains: physical data generation from sensors and observers, information transmission and storage, and cognitive data processing. 
This interconnected network supports various military functions like troop command, weapon systems management, surveillance, and medical care. However, ensuring robust device and network security remains a significant challenge, along with addressing the battlefield's connectivity, interoperability, and power requirements. Overcoming these hurdles is crucial for the future of military operations and underscores the importance of BLOS communication.

\section{Conclusion}
This paper explores BLOS communication's crucial role in defense, particularly where traditional Line-of-Sight methods fall short. To underscore BLOS capabilities, we've examined advanced technologies like phantom networks, nanonetworks, aerial relays, and satellite defense communication. The practical UAV path planning case study in artillery environments solidifies these concepts, demonstrating the tangible impact of BLOS defense systems in real scenarios. Moreover, the paper also highlights several future research directions, including resilience enhancement, the integration of heterogeneous networks, management of contested spectrums, advancements in multimedia communication, adaptive methodologies, and the burgeoning domain of the Internet of Military Things (IoMT).
In short, this research holds significance for academia and industry, opening paths for optimizing BLOS communication systems and guiding technological advancements with a focus on security and cyber resilience. It showcases practical BLOS applications in military operations, encouraging collaborations among defense agencies, tech developers, and academia. This resonates with the Sustainable Development Goals, fostering peace, security, and technological innovation while bridging academic and industrial frontiers.


\bibliographystyle{IEEEtran}
\bibliography{mybibliography}

\end{document}